\documentclass[prl,aps,twocolumn,longbibliography]{revtex4-2}

\usepackage[paperwidth=210mm,paperheight=297mm,centering,hmargin=1.63cm,vmargin=2.2cm]{geometry}

\usepackage{capt-of}

\pdfoutput=1

\newcommand{\kB}{k_{\textrm{B}}}

\newcommand{\Tc}{T_{\textrm c}}

\usepackage{float}
\usepackage{fourier}
\usepackage{amsmath,amssymb}
\usepackage{mathtools}
\usepackage{amsthm}
\usepackage{empheq}
\usepackage{cases}
\usepackage{times}
\usepackage{upgreek}
\usepackage{psfrag} 
\usepackage{latexsym} 
\usepackage{amstext}
\usepackage{amsfonts}
\usepackage{amsxtra} 
\usepackage{dcolumn}
\usepackage{dcolumn}
\usepackage{textcomp}
\usepackage{graphicx}
\usepackage{bm}
\usepackage{braket}
\usepackage{lineno}
\usepackage[svgnames,dvipsnames]{xcolor}
\usepackage{booktabs}
\usepackage[normalem]{ulem}
\usepackage{siunitx}

\definecolor{myColor}{rgb}{0.02,0.12,0.3}
\definecolor{myciteColor}{rgb}{0.39,0.7,0.89}
\usepackage[colorlinks=true,citecolor=myColor,linkcolor=myColor,urlcolor=myColor]{hyperref}

\def\be{\begin{equation}}
\def\ee{\end{equation}}

\def\nobreakbefore{%
  \relax\ifvmode\else
    \ifhmode
      \ifdim\lastskip > 0pt\relax
        \unskip\nobreakspace
      \else 
        \nobreakspace
      \fi
    \fi
  \fi
}
\let\oldcite\cite
\renewcommand\cite{\nobreakbefore\oldcite}

\makeatletter
\def\@fnsymbol#1{\ensuremath{\ifcase#1\or *\or \dagger\or \ddagger\or
   \mathsection\or \mathparagraph\or \|\or **\or \dagger\dagger
   \or \ddagger\ddagger \else\@ctrerr\fi}}
\makeatother

\begin{document} 
 
\title{
Observation of Vinen turbulence during far-from-equilibrium Bose--Einstein condensation\\
}
\author{Sebastian~J.~Morris$^{\ast}$}
\author{Martin~Gazo}
\author{Simon~M.~Fischer}
\author{Haoyu~Zhang}
\author{Christopher~J.~Ho$^{\dag}$}
\author{Nigel~R.~Cooper}
\author{Christoph~Eigen}
\author{Zoran~Hadzibabic}
\affiliation{
Cavendish Laboratory, University of Cambridge, J. J. Thomson Avenue, Cambridge CB3 0US, United Kingdom}
\date{\today}

\begin{abstract}
{Relaxation of far-from-equilibrium quantum fluids, intimately related to the emergence of long-range order, is theoretically associated with the decay of a turbulent isotropic tangle of vortex lines.}
We observe and study such decaying quantum turbulence in a homogeneous 3D atomic Bose gas. Using matter-wave techniques to magnify the gas density distribution, and then imaging a thin slice of the magnified cloud, we observe imprints of randomly oriented vortex lines and measure the vortex line-length density $\mathcal{L}$. The observed decay of $\mathcal{L}$ agrees with the prediction for Vinen `ultraquantum' turbulence. 
Although our weakly interacting gases are highly compressible, their large-scale dynamics are consistent with the behavior of an incompressible hydrodynamic fluid, with the decay of $\mathcal{L}$ not depending on the strength of the interatomic interactions and being similar to that in the strongly interacting superfluid helium.
\end{abstract}
\maketitle

A striking feature of quantum fluids is their irrotational flow, which leads to a quantized circulation that establishes topological defects in the form of vortex lines~\cite{Landau:1941,Onsager:1949,Feynman:1955,Svistunov:2015}.
The interactions and reconnections of quantized vortices are at the heart of a rich phenomenology of far-from-equilibrium dynamics, relevant across many fields and lengthscales~\cite{Vinen:1957,Vinen:1957c,Anderson:1975,Kibble:1976,Zurek:1985,Blatter:1994,Bulgac:2011}.
Central to these dynamics is the phenomenon of quantum turbulence, which has long been a cornerstone of research in superfluid helium~\cite{Tough:1982,Vinen:2002,Vinen:2010,Skrbek:2021,Barenghi:2023b}.

\begin{figure}[b!]
\centerline{\includegraphics[width=1\columnwidth]{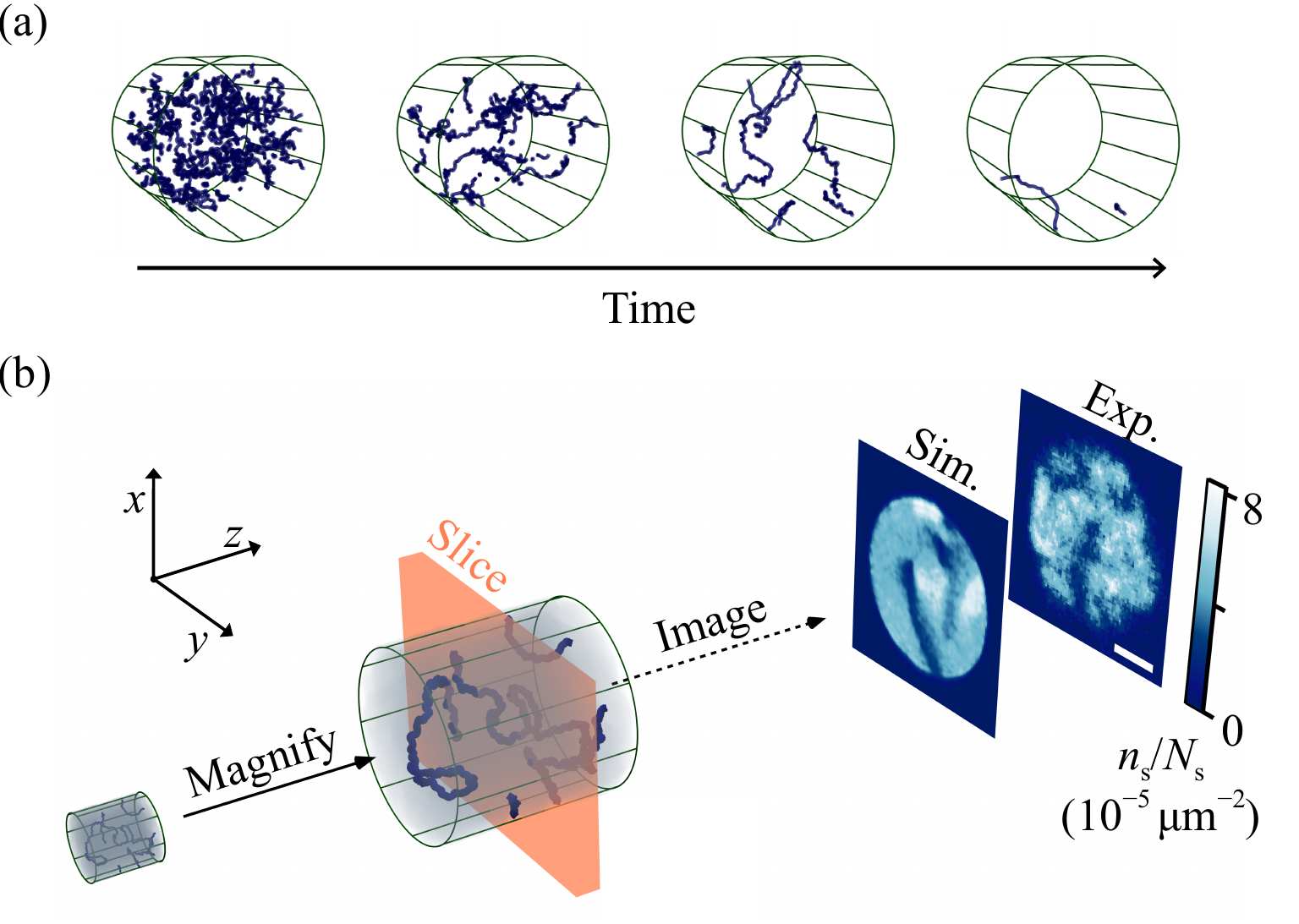}}
\caption{
Detecting decaying quantum turbulence during far-from-equilibrium condensation in an isolated gas.
(a) Cartoon of the expected vortex-tangle decay; the cylindrical container shows the geometry of our optical box trap~\cite{Gaunt:2013, Eigen:2016}.
(b) Experimental concept. We magnify the cloud~\cite{Murthy:2014,Asteria:2021,Brandstetter:2025} (by a factor of $\approx 3.5$ in the $x$-$y$ plane) and then image just a slice of it (orange shading)~\cite{Andrews:1997a}, so the imprints of the random vortex lines (dark lines) start and stop as they enter and leave the imaged slice.
We show an experimental image and a numerical simulation with similar parameters. Here $n_{\rm s}$ is the imaged 2D density, $N_{\rm s}$ is the total number of imaged atoms,  and the scale bar shows $50\,\upmu$m.
Our box trap has volume $V\approx 9 \times 10^4\,\upmu$m$^3$, the gas has density $n \approx 2.2 \,\upmu$m$^{-3}$, and the \emph{in-situ} vortex-core size is set by the healing length $\xi \sim 1\,\upmu$m; theoretically, the FWHM of the density dip associated with a vortex is $\approx 3\,\xi$~\cite{Fetter:2001}. The thickness of the imaged slice and the apparent vortex-line diameter are both $\approx 15\,\upmu$m; the latter is consistent with our magnification procedure and imaging resolution of $\approx 5\,\mu$m.
}
\label{fig1}
\end{figure}

The regime of Vinen or ultraquantum turbulence~\cite{Vinen:2002,Barenghi:2023b} corresponds to the decay of a random vortex tangle [see Fig.~\ref{fig1}(a)] without any large-scale (classical) flow.
Such turbulence has been theoretically linked to the dynamics of Kibble's strings in cosmology~\cite{Kibble:1976} and predicted to play a key role in far-from-equilibrium Bose--Einstein condensation (the emergence of long-range coherence)~\cite{Svistunov:1991,Svistunov:1995,Berloff:2002,Nowak:2012,Stagg:2016,Villois:2016,Noel:2025}.
Quantitatively, it is characterized by the vortex line-length density, $\mathcal{L}$, i.e.,~the total length of vortex lines per unit volume.
In liquid helium, the decay of $\mathcal{L}$ has been measured indirectly, using sound attenuation, already 70 years ago~\cite{Vinen:1957,Vinen:1957c}.
In this case, up to possible logarithmic corrections,
\begin{equation}
    \frac{\mathrm{d}\mathcal{L}}{\mathrm{d}t} = -B\mathcal{L}^2 \, ,
    \label{eq:QT}
\end{equation}
where $B$ is a constant with units of the quantum of circulation, $2\pi \hbar/m$, with $m$ being the particle mass~\cite{Svistunov:1995,Vinen:2002}.
Individual vortex lines have also been visualized using tracer particles, but such tracers are large compared to the vortex core and can modify the vortex dynamics~\cite{Bewley:2006,Paoletti:2008,Tang:2023,Peretti:2023}.

Ultracold atomic gases provide an excellent setting for studies of both driven and freely decaying turbulence~\cite{Henn:2009,Thompson:2014,Kwon:2014,Navon:2016,Seo:2017, Gauthier:2019,Johnstone:2019,Navon:2019,Liu:2021b,Galka:2022,Dogra:2023,Kim:2024,Karailiev:2024,Zhao:2025,Gazo:2025,Martirosyan:2025,Jiang:2026}.
In many situations, these systems also allow direct imaging of vortices~\cite{Matthews:1999, Madison:2000a, AboShaeer:2001, Bretin:2003b, Zwierlein:2005, Weiler:2008, Henn:2009, Freilich:2010, Neely:2010, Kwon:2014, Ku:2014, Donadello:2014,Serafini:2015, Seo:2017, Gauthier:2019, Johnstone:2019, Liu:2021b,Kwon:2021,Fletcher:2021,Kim:2024}, but a quantitative study of isotropic 3D vortex tangles has remained a challenge.
Recently, a study of far-from-equilibrium condensation in a homogeneous gas~\cite{Martirosyan:2025} revealed dynamics of the coherence length $\ell$ consistent with Eq.~(\ref{eq:QT}), assuming that $\ell$ is set by the typical vortex-line separation~\cite{Vinen:2002}, so $\ell \propto \mathcal{L}^{-1/2}$. This study was based on measurements of the gas momentum distribution, which is sensitive to the presence of vortices but does not directly reveal them.

In this Letter, we directly observe Vinen turbulence, during far-from-equilibrium condensation in a homogeneous 3D atomic gas.
To reveal imprints of the randomly oriented vortex lines, we magnify the cloud~\cite{Murthy:2014,Asteria:2021,Brandstetter:2025} and then image just a thin slice of it~\cite{Andrews:1997a}; see Fig.~\ref{fig1}(b).
We measure the vortex line-length density $\mathcal{L}$, observe a nonlinear decay of $\mathcal{L}$ in accordance with Eq.~(\ref{eq:QT}), and show that the prefactor $B$ does not depend on the strength of the (weak) interparticle interactions.
As in the strongly interacting superfluid helium~\cite{Walmsley:2008}, we find $B \approx \hbar/m$.

In our experiments, we prepare an isolated Bose gas in a far-from-equilibrium initial state by starting with a Bose--Einstein condensate and then perturbing it, as in Refs.~\cite{Martirosyan:2024,Martirosyan:2025}.
Our $^{39}$K gas is in the lowest hyperfine state, trapped in a cylindrical optical box~\cite{Gaunt:2013,Eigen:2016} of volume $V\approx 9 \times 10^4\,\upmu$m$^3$ (diameter $47(2)\,\upmu$m and length $52(2)\,\upmu$m), and has density $n \approx 2.2 \,\upmu$m$^{-3}$, corresponding to the critical temperature for condensation $\Tc \approx 70\,$nK. 
The initial state is incoherent but low energy, with energy per particle $\approx \kB\times 20\,\mathrm{nK}$, corresponding to the equilibrium condensed fraction $\eta \approx 0.5$~\cite{Martirosyan:2025}. We tune the interactions in the gas, characterized by the $s$-wave scattering length $a$, using the Feshbach resonance at $402.7$\,G~\cite{Etrych:2023}; the initial incoherent state is prepared at $a=0$ and we initiate the gas relaxation by turning on the interactions, at time $t=0$. In the interacting gas, the (equilibrium) radius of the vortex-line core is set by the healing length, $\xi = 1/\sqrt{8\pi na}$, of order $1\,\upmu$m.

To observe vortex imprints, we switch off the box trap and the interactions ($a \to 0$), and magnify the cloud~\cite{Murthy:2014,Asteria:2021,Brandstetter:2025} in the $x$-$y$ plane by a factor $M \approx 3.5$ [see Fig.~\ref{fig1}(b)]; as in Refs.~\cite{Shvarchuck:2002,Tung:2010,Murthy:2014,Fletcher:2015,Asteria:2021,Brandstetter:2025}, we use a pulsed harmonic potential as a matter-wave lens~\footnote{We use a red-detuned laser beam to realize a harmonic trap, with frequencies $\omega_x \approx 2 \pi \times 40\,$Hz, $\omega_y \approx 2 \pi \times 45\,$Hz, and $\omega_z \approx 0$. The slight anisotropy in the $x$-$y$ plane limits our resolution to a distance that corresponds to $\approx 2\,\upmu$m \emph{in-situ}.}.
We then image just a central slice of the cloud~\cite{Andrews:1997a}, of thickness $d\approx15\,\upmu$m~\footnote{We use a laser beam shaped by a digital micromirror device to optically pump only a slice of the cloud into the highest hyperfine ground state, and then image only the atoms in that state.}. 
The vortex lines that cross this slice appear as high-contrast dips (dark lines) in the imaged 2D density $n_{\rm s}$.
Theoretically, the {\it in-situ} FWHM of the density depletion due to a vortex core is $\approx 3\,\xi$~\cite{Fetter:2001}, while  the apparent vortex-core diameter in our images ($\approx 15\,\upmu$m) is slightly larger than $3M\xi$, both due to our imaging resolution ($\approx 5\,\upmu$m) and because the free expansion along $z$ during the $19\,$ms of $x$-$y$ magnification leads to some blurring.

\begin{figure}[t!]
\centerline{\includegraphics[width=1\columnwidth]{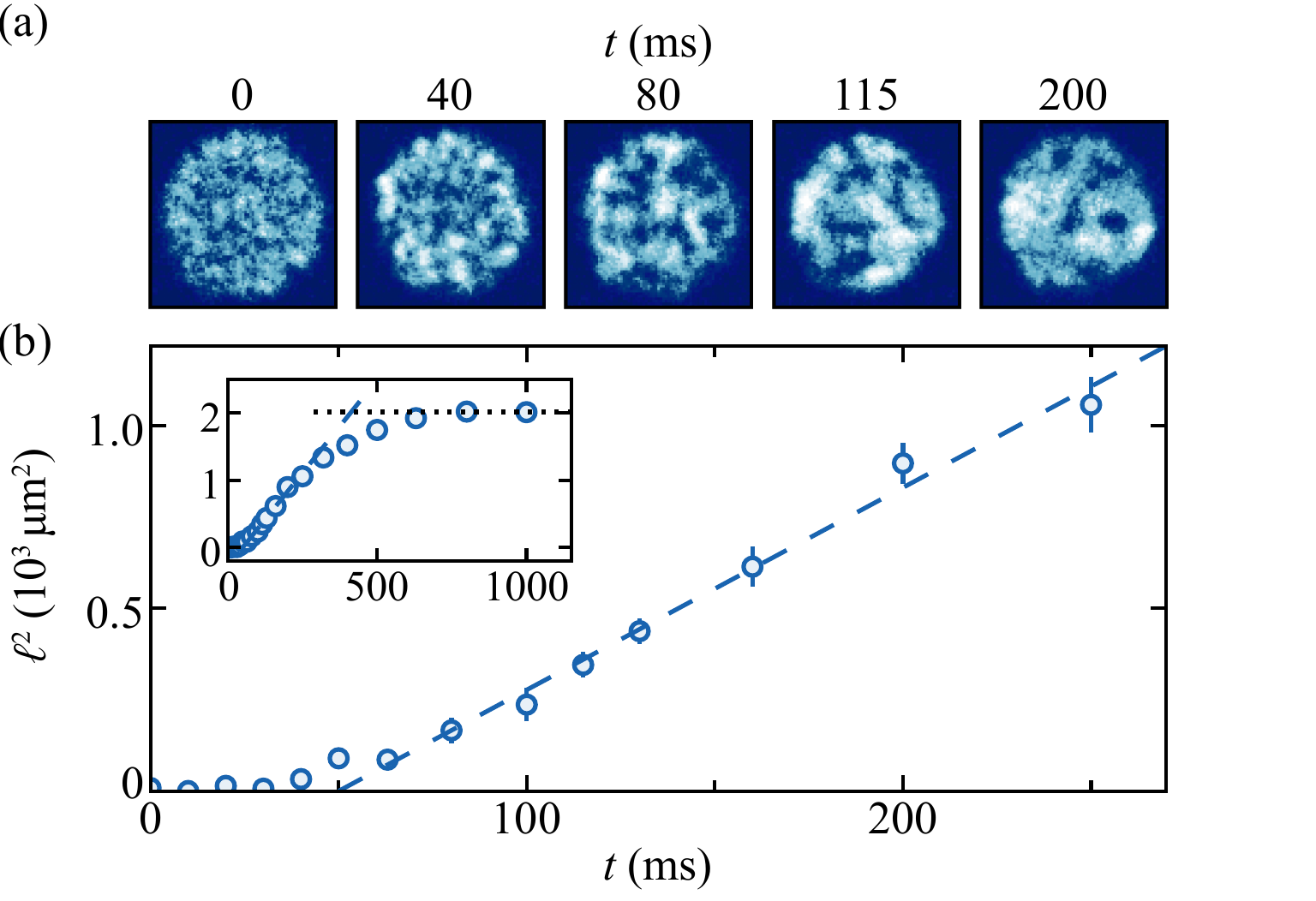}}
\caption{
Stages of relaxation, starting in an incoherent state. Here the scattering length is $a=430\,a_0$, so $\xi = \SI{0.9}{\um}$.
(a) Examples of cloud slices at different relaxation times, imaged in separate experimental runs; the scale bar and the color scale are the same as in Fig.~\ref{fig1}(b). The early-time images show only small-scale density fluctuations, whereas for $t \gtrsim 80\,$ms we observe well-defined and well-separated vortex lines.
(b) The corresponding evolution of the coherence length $\ell$, obtained separately from measurements of momentum distributions, as in Ref.~\cite{Martirosyan:2025}. We find that the onset of universal coarsening seen in $\ell^2(t)$, given by ${\rm{d}} \ell^2/{\rm{d}}t \approx 3.4 \,\hbar / m$ (dashed line)~\cite{Martirosyan:2025}, and the emergence of well-defined vortices essentially coincide. The inset shows how $\ell^2(t)$ saturates at longer times; the dotted line shows $V^{2/3}$.}
\label{fig2}
\end{figure}

In Fig.~\ref{fig2}(a), we show characteristic slice images for different relaxation times, with $a$ set to $430\,a_0$ (where $a_0$ is the Bohr radius). At early times we observe only small-scale density fluctuations, while at long times (here $t\gtrsim 80\,$ms) we observe well-defined and well-separated vortex imprints.

For comparison, in Fig.~\ref{fig2}(b) we show the corresponding evolution of $\ell^2(t)$, which shows universal coarsening, given by ${\rm{d}} \ell^2/{\rm{d}}t \approx 3.4 \,\hbar / m$~\cite{Martirosyan:2025}, once $\ell$ significantly exceeds $\xi$ and until it becomes comparable to the system size; as in Ref.~\cite{Martirosyan:2025}, we obtain $\ell(t)$ from momentum-distribution measurements and normalize it such that $\ell^3 = V$ for an equilibrium condensate.

This comparison shows that the emergence of well-defined vortices essentially coincides with the onset of the universal coarsening seen in $\ell^2(t)$. Such coarsening  does not directly reveal vortices, but is indeed consistent with the Vinen scenario of well-separated vortex lines, with $\ell^2 \sim 1/\mathcal{L}$~\cite{Vinen:2002,Barenghi:2023b}. It is also consistent with the notion that at lengthscales $\gg \xi$ the gas is effectively incompressible~\cite{Kraichnan:1967}, similar to the effective incompressibility of classical gases for Mach numbers $\ll 1$~\cite{Tritton:2012}.

\begin{figure*}[b!]
\centerline{\includegraphics[width=1\textwidth]{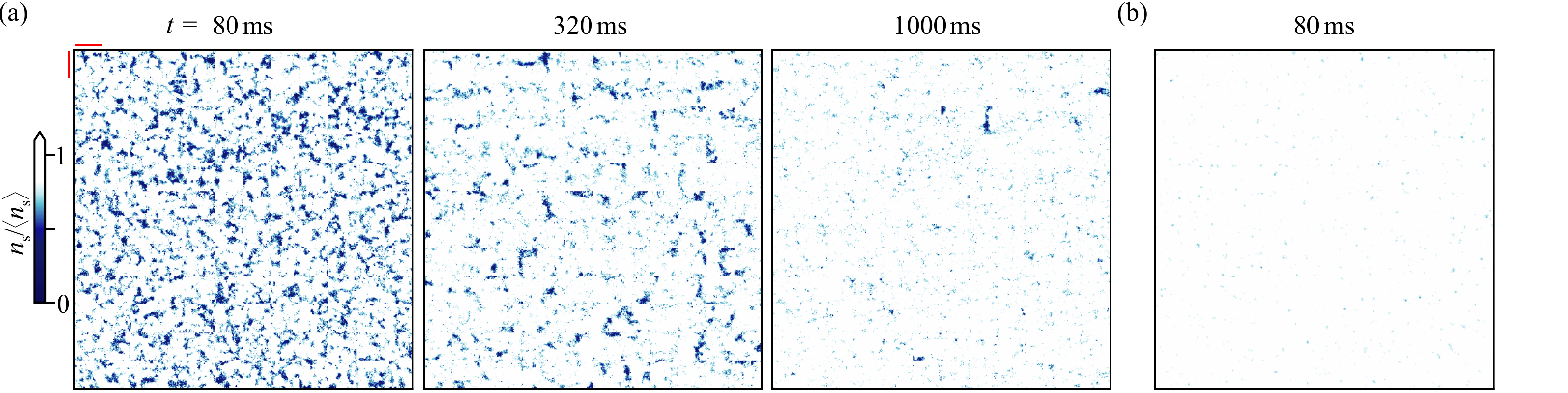}}
\caption{
Visualizing the decay of vortex lines.
(a) For $t \ge 80\,$ms, when vortices are well defined (see Fig.~\ref{fig2}), we tile cropped slice images corresponding to $12\times 12$ consecutive experimental realizations, which show randomly varying vortex configurations.
The red bars in the top-left corner indicate the size of a single-image tile, which corresponds to $\approx 30\,\upmu{\rm m} \times 30\,\upmu{\rm m}$ \emph{in situ}, and $\langle n_{\rm s}\rangle$ is the average over the experimental repetitions.
(b) Here, for $t=80\,$ms, we tile images taken without slicing (i.e., with line-of-sight integration over the whole magnified cloud). In this case, with the same color scale as in (a), no vortex imprints are observed. 
}
\label{fig3}
\end{figure*}

The long-time vortex configurations for fixed $t$ vary randomly between experimental realizations, but statistically the number of vortices decays.
We illustrate this in Fig.~\ref{fig3}(a) for $t \geq 80\,$ms.
Here, for each $t$, we tile $12 \times 12$ consecutively taken images, cropped to the central square area corresponding to $\approx 30\,\upmu{\rm m}\times 30\,\upmu{\rm m}$ {\it in situ} ($\approx 100\,\upmu{\rm m}\times 100\,\upmu{\rm m}$ after magnification).
At times significantly beyond $t=320\,$ms, vortex imprints appear in a small fraction of experimental realizations; as our tiling visually suggests, in a much larger system their typical separation would be larger than our system size.
For 

\onecolumngrid
~
\vspace{-2em}
\clearpage
\begin{figure*}[t!]
\centerline{\includegraphics[width=1\textwidth]{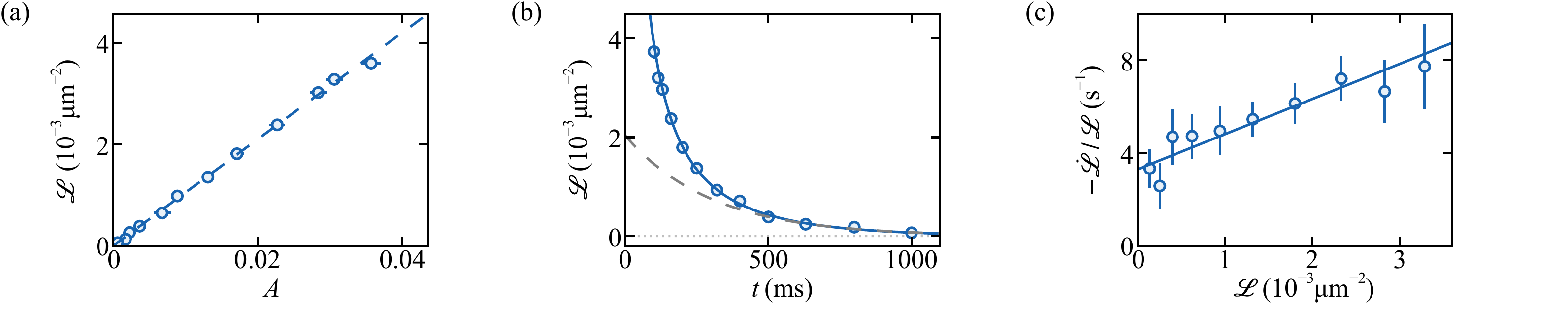}}
\caption{
Quantifying the decay of vortex lines.
(a)~The vortex line-length density $\mathcal{L}$, extracted according to Eq.~(\ref{eq:Ni}), is simply proportional to $A$, the average fractional area of images covered by the vortex imprints [see Fig.~\ref{fig3}(a)].
(b)~The evolution of $\mathcal{L}$. The dashed line shows an exponential fit, corresponding to one-body vortex decay, to data with $t\geq 500\,$ms.
The faster initial decay hints at vortex-vortex interactions.
The solid line shows a fit based on Eq.~(\ref{eq:QT3}), with $B=1.0(2) \, \hbar/m$ and $\tau = 300(40)\,$ms. 
(c)~By extracting $\dot{\mathcal{L}} \equiv {\rm{d}} \mathcal{L}/{\rm{d}}t$ at different times during the relaxation, we explicitly recover the differential Eq.~(\ref{eq:QT2}), with consistent $B=0.9(2)\, \hbar/m$ and $\tau = 300(50)\,$ms, respectively the slope and inverse intercept of the solid line.
}
\label{fig4}
\vspace{-1em}
\end{figure*}
\twocolumngrid

\noindent
comparison, in Fig.~\ref{fig3}(b) we show that if we image the whole cloud, instead of just a slice, no vortex imprints are observed.

\begin{figure*}[t!]
\centerline{\includegraphics[width=1\textwidth]{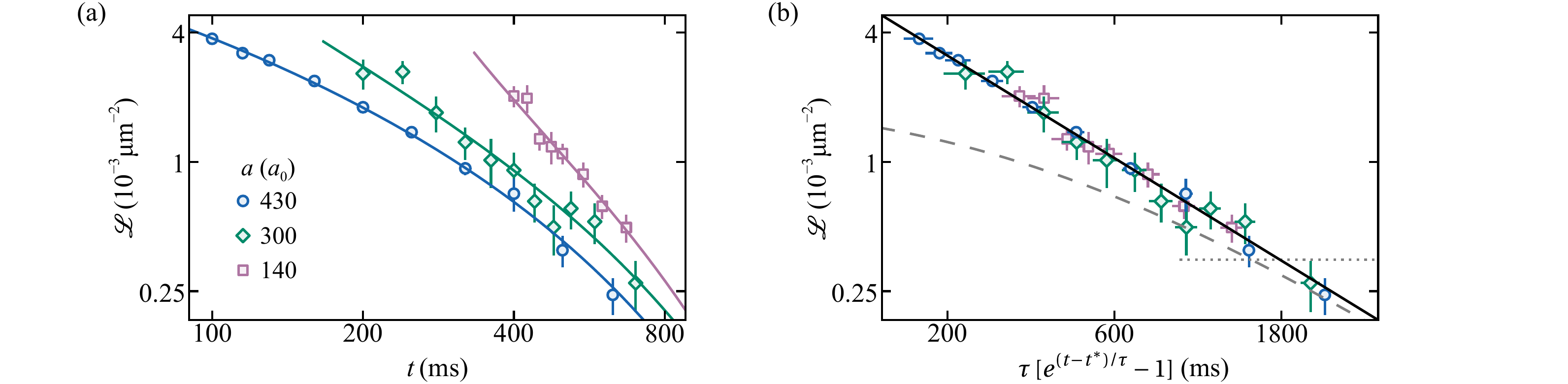}}
\caption{
Universality of Vinen turbulence, for varying $a$ (set at $t=0$).
(a)~The evolution of $\mathcal{L}$ for different $a$ is always fit well (solid lines) by Eq.~(\ref{eq:QT3}), with only the non-universal $t^*$ depending on $a$.
Here, all three data sets are fit with the same $B=1.0(1)\, \hbar/m$; fitting them independently gives consistent $B$ values.
Accounting for the difference in the particle mass, this value of $B$ is essentially the same as obtained for superfluid helium~\cite{Walmsley:2008}.
(b)~Plotting the same data according to Eq.~(\ref{eq:QT3}), with $a$-dependent $t^*$, collapses them onto the same universal curve; the solid line corresponds to $B=1.0\, \hbar/m$.
The dashed line shows the exponential with timescale $300\,$ms that coincides with the solid line at long times.
The horizontal dotted line, $\mathcal{L} = 3.5 \times 10^{-4}\,\upmu$m$^{-2}$, corresponds to (on average) just one vortex line being present in our whole box trap; at smaller $\mathcal{L}$, the dynamics are captured by the one-body exponential decay.
}
\label{fig5}
\end{figure*}

For the quantitative analysis of our data, we employ the general mathematical result for an isotropic random tangle of lines~\cite{Binzoni:2025},
\begin{equation}
    \mathcal{L} = 4N_{\rm i}/S \, ,
    \label{eq:Ni}
\end{equation}
where $N_{\rm i}$ is the number of intersections of lines with a subvolume of surface area $S$.
In our case, $N_{\rm i}$ is the number of distinct vortex imprints, and $S$ is the \emph{in-situ} surface area corresponding to the region where we count them~\footnote{For the data in Figs.~\ref{fig4} and~\ref{fig5}, we count imprints in our images within a central circle of radius $r_{\rm D}\approx 52\,\upmu$m, which corresponds to \emph{in-situ} $S \approx 2800\,\upmu$m$^2$.
None of our results change (within errorbars) if we reduce $r_{\rm D}$ to $\approx 34\,\upmu{\rm m}$, which corresponds to $S \approx 1500\,\upmu$m$^2$.
}.  

We define the image area covered by the imprints by $n_{\rm s} < \alpha \langle n_{\rm s} \rangle$, where $\langle n_{\rm s} \rangle$ is the average $n_{\rm s}$, and set the cutoff $\alpha = 0.5$ to capture the high-contrast depletions associated with the vortices while excluding other density variations (including noise). We then count $N_{\rm i}$ by grouping the imprint pixels into the same cluster when they are separated by $\leq 4 \,{\rm pixels} \approx 10\,\upmu{\rm m}$~\footnote{We exclude clusters with fewer pixels than $N_{\rm min} = 4$, which is sufficiently large to avoid counting noise, so $N_{\rm i}$ vanishes for $t\rightarrow \infty$.}.

More intuitively, $\mathcal{L}$ is proportional to the average fractional area, $A$, covered by the imprints.
We show in Fig.~\ref{fig4}(a) that $\mathcal{L}$ extracted according to Eq.~(\ref{eq:Ni}) is indeed proportional to $A$.
However, Eq.~(\ref{eq:Ni}) provides an absolute calibration of $\mathcal{L}$ in a way that does not depend on the vortex-core size and allows direct comparison of the dynamics at different~$a$.

In Fig.~\ref{fig4}(b), we plot the extracted $\mathcal{L}(t)$ for $a=430\,a_0$ and $t> 80\,$ms.
At very long times (very small $\mathcal{L}$), the data are described well by a simple exponential decay (dashed line), with a time constant $\approx 300\,$ms; such one-body decay is suggestive of individual vortices annihilating on the trap walls with their images~\cite{Schwarz:1985}.
However, at earlier times we observe a decay faster than the exponential one, which we attribute to the vortex-vortex interactions that drive Vinen turbulence.
Augmenting Eq.~(\ref{eq:QT}) to include one-body decay, we have:
\begin{equation}
  \frac{\mathrm{d}\mathcal{L}}{\mathrm{d}t} = -B\mathcal{L}^2  - \frac{\mathcal{L}}{\tau} \, , 
 \label{eq:QT2}
\end{equation}
which after integration gives
\begin{equation}
  \mathcal{L} =  \frac{1}{B\tau \left[ \exp \left( \frac{t - t^*}{\tau} \right) - 1   \right]} \, , 
 \label{eq:QT3}
\end{equation}
where $t^*$ is the integration constant.

The solid line in Fig.~\ref{fig4}(b) shows a fit based on Eq.~(\ref{eq:QT3}).
We find excellent agreement with the data, with $B=1.0(2)\, \hbar/m$.
Moreover, by numerically differentiating $\mathcal{L}(t)$, in Fig.~\ref{fig4}(c), we recover the differential Eq.~(\ref{eq:QT2}), with a consistent $B=0.9(2) \, \hbar/m$ (the slope of the solid line).

Finally, in Fig.~\ref{fig5}, we generalize our study to different $a$ (set at $t=0$).
In Fig.~\ref{fig5}(a), we show for three different $a$ values (the $430\,a_0$ data being the same as in Fig.~\ref{fig4}) that $\mathcal{L}(t)$ is always fit well using Eq.~(\ref{eq:QT3}) with the same $B=1.0(1)\, \hbar/m$.
Note that here the three curves are offset from each other because (only) the $t^*$ values are different for different $a$; for weaker interactions, the system takes longer to develop well-defined vortices, but then evolves in an interaction-independent way (see also~\cite{Martirosyan:2025}).

In Fig.~\ref{fig5}(b), we collapse all the data from Fig.~\ref{fig5}(a) by plotting them according to Eq.~(\ref{eq:QT3}) with $a$-dependent $t^*$; the solid line corresponds to $B=1.0\,\hbar/m$.
The dashed line shows an exponential with timescale $300\,$ms [as in Fig.~\ref{fig4}(b)], and the horizontal dotted line corresponds to (on average) just one vortex line being present in the whole box trap [see Eq.~(\ref{eq:Ni})].

Accounting for the difference in the particle mass, our $B$ is similar to that observed in superfluid $^4$He, where $B \approx \hbar/m_{\rm He}$~\cite{Walmsley:2008}, despite the large differences between the microscopics of the two systems (even in equilibrium) and the fact that in our case the typical ratio of the vortex-line separation, $\mathcal{L}^{-1/2}$, and the vortex-core size is smaller by about $5$ orders of magnitude.
Numerical simulations of the Gross--Pitaevskii equation also give $B \sim \hbar / m$~\cite{Stagg:2016,Villois:2016} (see also~\cite{Tsubota:2000,Baggaley:2012,Shukla:2019}).

In conclusion, we have observed random vortex lines during far-from-equilibrium Bose--Einstein condensation in an isolated homogeneous 3D gas, and shown their decay to be consistent with Vinen turbulence.
We observe well-defined vortices and their universal decay in the regime of gas relaxation when the coherence length is already much larger than the vortex core size; in the future, it would be interesting to also study the earlier stages of relaxation and the expected transition from wave to vortex-dominated turbulence~\cite{Svistunov:2015}.
It would also be interesting to increase the energy per particle and explore vortex-relaxation dynamics close to criticality.
Finally, imaging multiple slices of the same cloud would allow full tomography of vortex tangles and studies of their structural dynamics.


We thank Gevorg Martirosyan, Ji\v r\'i Etrych, Nikolai Maslov, Carlo Barenghi, and Mike Gunn for discussions.
This work was supported by ERC [UniFlat], EPSRC [Grant No.~EP/Y01510X/1], and STFC [Grants No.~ST/T006056/1 and No.~ST/Y004469/1].
Z.~H. acknowledges support from the Royal Society Wolfson Fellowship.



%

\clearpage
\setcounter{figure}{0} 
\setcounter{equation}{0}

\renewcommand\theequation{S\arabic{equation}} 
\renewcommand\thefigure{\arabic{figure}} 
\renewcommand{\figurename}[1]{FIG.~S}

\end{document}